\documentclass[twocolumn,prb,showpacs]{revtex4}
\usepackage{graphicx}
\usepackage{epsfig}

\makeatother
\begin{document}

\title{Thermodynamic properties of tetrameric bond-alternating spin chains}

\author{H.T. Lu,$^{1,2}$ Y.H. Su,$^{3}$ L.Q. Sun,$^{2}$ J. Chang,$^{2}$
C.S. Liu,$^{2}$ H.G. Luo,$^{2}$ and T. Xiang$^{2}$}

\affiliation{${}^{1}$School of Physics, Peking University, Beijing 100871, China\\
${}^{2}$Institute of Theoretical Physics and Interdisciplinary Center
of Theoretical Studies, Chinese Academy of Sciences, Beijing 100080, China \\
${}^{3}$Center for Advanced Study, Tsinghua University, Beijing 100084,China}

\date{Received 17 December 2004}

\begin{abstract}
Thermodynamic properties of a tetrameric bond-alternating
Heisenberg spin chain with
ferromagnetic-ferromagnetic-antiferromagnetic-antiferromagnetic
exchange interactions are studied using the transfer-matrix
renormalization group and compared to experimental measurements.
The temperature dependence of the uniform susceptibility exhibits
typical ferrimagnetic features. Both the uniform and staggered
magnetic susceptibilities diverge in the limit $T\rightarrow 0$,
indicating that the ground state has both ferromagnetic and
antiferromagnetic long-range orders. A double-peak structure
appears in the temperature dependence of the specific heat. Our
numerical calculation gives a good account for the temperature and
field dependence of the susceptibility, the magnetization, and the
specific heat for Cu(3-Clpy)$_{2}$(N$_{3}$)$_{2}$
(3-Clpy=3-Chloroyridine).
\end{abstract}


\maketitle

\section{Introduction}

Physical properties of one-dimensional quantum ferrimagnets have
attracted much interest in recent years. In particular, physical
properties of the so-called Lieb-Mattis\cite{Lieb}-type
ferrimagnets have been extensively investigated both theoretically
and experimentally.\cite{Pati,Yamamoto98prb,yamamoto98,wu} In
these materials, there exists strong competition between
ferromagnetic and antiferromagnetic fluctuations. This competition
leads to a ferrimagnetic-ordered ground state with coexisting
gapless ferromagnetic and gapped antiferromagnetic excitations, a
minimum in the temperature dependence of the product of the
susceptibility and temperature, a double-peak structure in the
specific heat, and a wide variety of other physical phenomena.

A Lieb-Mattis-type ferrimagnet can be constructed from monospin
chains with polymerized exchange interactions. It can be also
constructed from an alternating spin chain with antiferromagnetic
interactions. These two types of ferrimagnets are equivalent to
each other in certain limits. Many of their thermodynamic or
dynamic response functions show qualitatively similar behaviors.
For convenience in the discussion below, we call the ferrimagnets
formed from monospins or from alternating large and small spins as
Type-I or Type-II ferrimagnets.

The simplest bond-alternating spin chain is a
ferromagnetic-antiferromagnetic (F-AF) alternating Heisenberg spin
chain. A typical example of this F-AF alternating
spin chain is the compound (CH$_{3}$)$_{2}$CHNH$_{3}$CuCl$_{3}$.
Manaka \emph{et al.}\cite{Manaka} did birefringence measurement on
this material. Although the ground state of the compound is
nondegenerate, a double-peak structure of the specific heat was
observed.

A Type-I ferrimagnet can be constructed from a tetrameric
ferromagnetic-ferromagnetic-antiferromagnetic-antiferromagnetic
(F-F-AF-AF) Heisenberg exchange spin chain. Recently, Hagiwara
\emph{et al}. studied thermodynamic
properties\cite{Hagiwara1,Hagiwara2} of a tetrameric chain
compound Cu(3-Clpy)$_{2}$(N$_{3}$)$_{2}$(3-Clpy=3-Chloroyridine),
abbreviated as CCPA below. \cite{Inorg} Typical behavior of
ferrimagnets was revealed. Furthermore, by exactly diagonalizing
an isotropic F-F-AF-AF Heisenberg model on small lattice, they
found that the experimental data for the magnetization and
susceptibility can be quantitatively understood from this model.
\cite{Hagiwara1,Hagiwara2} There were also other theoretical and
numerical studies on this tetrameric chain system. Yamamoto
calculated the zero-field specific heat and susceptibility of the
F-F-AF-AF Heisenberg model using the modified spin-wave theory as
well as the Quantum Monte Carlo (QMC) method.\cite{yamamoto04} Nakanishi and
Yamamoto\cite{Nakanishi} found that the specific heat of this
model shows a double-peak structure. However, a direct comparison
between experiments and numerical calculations, especially for the
temperature dependence of the specific heat, is still absent.

A typical Type-II ferrimagnet is the alternating $S=1$ and $1/2$
spin chain. Kahn's group\cite{kahn} synthesized successfully a
class of bimetallic chain compounds with each unit cell containing
two spins with different values. Typical compounds include
ACu(pba)(H$_{2}$O)$_{3}\cdot n$H$_{2}$O and
ACu(pbaOH)(H$_{2}$O)$_{3}\cdot n$H$_{2}$O, where A = Mn, Fe, Co,
Ni, Zn, pba = 1,3-propylenebis(oxamato), and pbaOH = 2-hydroxy-1,
3-propylenebis(oxamato). This has stimulated theoretical studies
on alternating spin chains. From numerical simulations, Pati
\emph{et al.}\cite{Pati} revealed the coexistence of gapless and
gapped excitations in alternating spin-$(1,1/2)$ ferrimagnetic
chains using the density-matrix renormalization group (DMRG).\cite{White} The
coexistence of ferromagnetic and antiferromagnetic aspects was also
found by Brehmer \emph{et al}.\cite{brehmer} from 
the Quantum Monte Carlo (QMC) simulations. Thermodynamic properties of the
alternating spin chains were studied by the
transfer-matrix renormalization group (TMRG).\cite{yamamoto98} To
further explore the dual features of ferrimagnetism, the Schwinger
boson and the spin-wave approximations were exploited.
\cite{wu,yamamoto98,yamamoto04}

In this paper, we investigate numerically thermodynamic properties
of tetrameric spin chains with alternating F-F-AF-AF exchange
interactions and compare the results to the experimental data
for CCPA. In Sec. \ref{sec:model}, the isotropic F-F-AF-AF
Heisenberg exchange model for the tetrameric spin chains with its
general properties is introduced. In Sec.
\ref{sec:Main-Numerical-Results}, thermodynamic quantities of the
tetrameric model including the magnetization, uniform and
staggered magnetic susceptibilities, and field dependent specific
heat, are evaluated using the transfer-matrix renormalization
group. \cite{Bursill,Wang,Xiang} Section
\ref{sec:Comparison-With-Experiments} compares the numerical
results to the experimental data on CCPA. Our numerical results
agree well with the experimental data for the zero-field uniform
susceptibility as well as the magnetization. The numerical results
for the difference of the specific heat between two different
fields are also in good agreement with the experimental data. We briefly 
discuss the sharp low-temperature peak that appeared in
the zero-field specific-heat curve published in Refs.
\onlinecite{Hagiwara1} and \onlinecite{Hagiwara2}. It is suggested that this experimental sharp
peak is an extrinsic feature and most probably because of the
contribution of magnetic impurities.

\section{Model Hamiltonian and general properties\label{sec:model}}

\begin{figure}
\includegraphics[width=7.5cm]{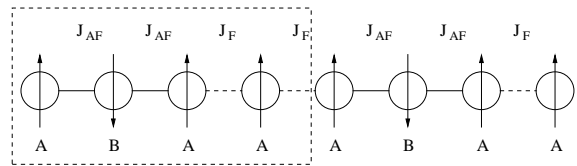}
\caption{\label{tetramerpic}\emph{Schematic} representation of the chain
structure of Cu(3-Clpy)$_{2}$(N$_{3}$)$_{2}$ (CCPA) with
interactions described by Hamiltonian (\ref{eq:1}). The system can
be decomposed into two sublattices according to the nature of
coupling, \emph{i.e.}, A and B sublattices, respectively.}
\end{figure}

CCPA is a tetrameric spin-chain compound. It consists of copper
chains in which adjacent Cu$^{2+}$ ions are linked by two kinds of
azido bridges, namely, end-on and end-to-end bridge, respectively
[Fig. 1 of Ref. \onlinecite{Hagiwara1}]. The exchange interactions
between adjacent Cu$^{2+}$ ions are determined by the azido
bridges linked to them. The low-energy magnetic excitations in this
material can be modeled by a $S=1/2$ Heisenberg Hamiltonian
defined by
\begin{eqnarray}
H & = & H_{0}+H^{\prime},\label{eq:1}\\
H_{0} & = & \sum_{i=1}^{L}\left[J_{F}\mathbf{S}_{4i-3}
\cdot\mathbf{S}_{4i-2}
+J_{AF}\mathbf{S}_{4i-2}\cdot\mathbf{S}_{4i-1} \right. \nonumber
\\
&&\left. +J_{AF} \mathbf{S}_{4i-1}\cdot\mathbf{S}_{4i}
+J_{F} \mathbf{S}_{4i}\cdot\mathbf{S}_{4i+1}\right], \\
H^{\prime} & = & -\sum_{i=1}^{4L}g\mu_{B}hS_{i}^{z},
\end{eqnarray}
where $L$ is the number of unit cells, $J_{F}$ and $J_{AF}$ are
the ferromagnetic and antiferromagnetic exchange constants,
respectively, $g$ is the magnetic $g$ factor and $h$ is the
external magnetic field. This model belongs to the family of
Lieb-Mattis-type ferrimagnets, as shown in Fig. \ref{tetramerpic}.
According to the Lieb-Mattis theorem,\cite{Lieb} this tetrameric
system has $(2L+1)$-fold degenerate ground states with total spin
$S_{\rm tot}=L$. The macroscopic magnetization of the ground state
has been quantitatively confirmed by magnetization measurements
with pulsed and static fields on a single crystal of
CCPA.\cite{Hagiwara2}

In the limit $J_{F}=0$, the system is decoupled into isolated
trimers, separated by local one-half spins. The Hamiltonian then
becomes
\begin{equation}
H=\sum_{i=1}^{L}H_{i}^{(c)},\label{eq:2}
\end{equation}
where
\begin{eqnarray}
H_{i}^{(c)} & = &
J_{AF}\left(\mathbf{S}_{4i-2}\cdot\mathbf{S}_{4i-1}
+\mathbf{S}_{4i-1}\cdot\mathbf{S}_{4i}\right) \nonumber \\
& &
-g\mu_{B}h\left(S_{4i-3}^{z}+S_{4i-2}^{z}+S_{4i-1}^{z}+S_{4i}^{z}\right).
\end{eqnarray}
This model is exactly soluble. In the absence of an applied field,
an antiferromagnetic trimer is a three-level system with a doubly
degenerate ground state of $S_{\rm tot}=1/2$. One excited state is
twofold degenerate with $S_{\rm tot}=1/2$ and an energy $J_{AF}$
above the ground state. The other is fourfold degenerate with
$S_{\rm tot}=3/2$ and an energy $3J_{AF}/2$ above the ground
state. For finite but small $J_{F}$, the ferromagnetic correlation
dominates the low-temperature excitations. The low temperature
behavior of the system is expected to act similarly as a
spin-$\frac{1}{2}$ ferromagnetic spin chain.

In the limit $\mid J_{F}\mid\gg J_{AF}$, the three
ferromagnetically coupled neighbors in one unit will bind together
to form a $3/2$ spin. In this case, the system is expected to
behave similarly as a $\left(S,s\right)=\left(3/2,1/2\right)$
Heisenberg ferrimagnetic spin chain.

For the alternating $\left(S,s\right)=\left(1,1/2\right)$
Heisenberg spin chain, \cite{Pati,brehmer,yamamoto98,wu} there
exist both gapless ferromagnetic and gapped antiferromagnetic
excitations. For the bond-alternating ferrimagnetic spin chain
considered here, these two kinds of excitations are also expected
to exist. In low temperatures, thermodynamic properties of the
system are more strongly affected by ferromagnetic excitations.
However, with increasing temperatures, the contribution from
gapped antiferromagnetic excitations increases. This leads to a
crossover in the behavior of $C$ and $\chi$. For instance, in low
temperatures, for the $\left(S,s\right)=\left(1,1/2\right)$ spin
chains, the specific heat and the susceptibility vary respectively
as $C\propto T^{1/2}$ and $\chi\propto T^{-2}$, same as for the
spin-$\frac{1}{2}$ ferromagnetic spin chain. However, in the intermediate
temperature regime, they show a Schottky-like peak and a minimum,
respectively. \cite{yamamoto98}

\section{Numerical Results\label{sec:Main-Numerical-Results}}

In this section, we will use the TMRG to evaluate the temperature
dependence of the magnetic susceptibility and the specific heat
for the tetrameric bond-alternating spin model described by
Hamiltonian (\ref{eq:1}). The TMRG is a powerful numerical tool
for studying thermodynamic properties of one-dimensional quantum
systems.\cite{Bursill,Wang,Xiang} It starts by expressing the
partition function of a one-dimensional quantum lattice system as
a trace of a virtual transfer matrix $\mathcal{T}_{M}$ using the
Trotter-Suzuki decomposition
\begin{equation}
Z=\mathrm{Tr}e^{-\beta
H}=\lim_{M\rightarrow\infty}\mathrm{Tr}\mathcal{T}_{M}^{N/2},
\end{equation}
where $M$ is the Trotter number, $N$ is the system size, and
$\tau=\beta/M$. For the tetrameric model considered here, $N= 2L$
and each site used in the Trotter-Suzuki decomposition contains
two spins. In the limit $N\rightarrow\infty$, one can evaluate
nearly all thermodynamic quantities by the maximum eigenvalue
$\lambda_{\textrm{max}}$ and the corresponding left
$\langle\psi^{L}\mid$ and right $\mid\psi^{R}\rangle$ eigenvectors
of the transfer matrix $\mathcal{T}_{M}$. For example, the free
energy $F$, internal energy $U$, and longitudinal magnetization $M_z$
per unit cell can be expressed, respectively, as
\begin{eqnarray}
F & = &-\lim_{N\rightarrow\infty}\frac{1}{N\beta}\ln Z
= -\frac{1}{2\beta}\lim_{M\rightarrow\infty}\ln\lambda_{\textrm{max}},\\
U & = &\frac {\langle\psi^{L}\mid\tilde{\mathcal{T}}_{U}
\mid\psi^{R}\rangle}{\lambda_{\textrm{max}}},\\
M_z & = & \frac{\langle\psi^{L}\mid\tilde{\mathcal{T}}_{M}
\mid\psi^{R}\rangle}{\lambda_{\textrm{max}}},
\end{eqnarray}
where the definition of the transfer matrices
$\tilde{\mathcal{T}}_{U}$ and $\tilde{\mathcal{T}}_{M}$ can be
found from Refs. \onlinecite{Wang} and \onlinecite{Xiang}. The specific heat and
magnetic susceptibility can then be calculated by numerical
derivatives of $U$ and $M_z$, respectively,
\begin{eqnarray}
C & =&\frac{\partial U}{\partial T},\\
\chi & =&\frac{\partial M_z}{\partial H}.
\end{eqnarray}

In our TMRG iterations, $60$ states are retained in the
calculation of the susceptibility, and $80$ states are retained in
the calculation of the specific heat. The error results from the
Trotter-Suzuki decomposition is less than $10^{-3}$. The
truncation errors are smaller than $10^{-8}$.

\begin{figure}
\includegraphics[width=7.5cm]{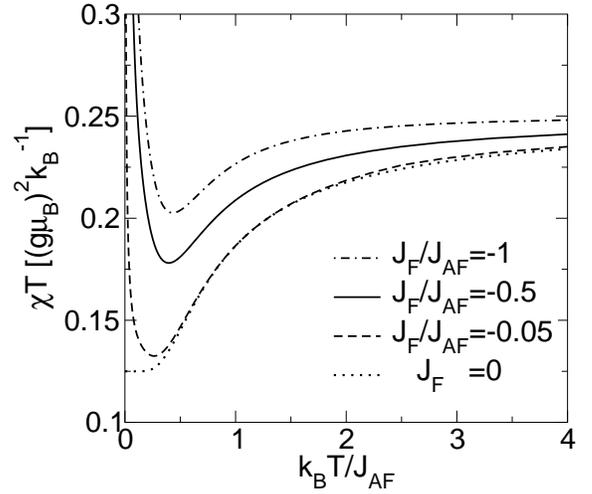}
\caption{\label{sus_ratio}TMRG results (except for the case
$J_{F}=0$) of the uniform zero-field susceptibility multiplied by
temperature for the tetrameric bond-alternating spin model.}
\vspace{0.8cm}
\end{figure}

Figure \ref{sus_ratio} shows the temperature dependence of the
uniform zero-field susceptibility multiplied by temperature $\chi
T$. When $J_{F}=0$, $\chi T$ approaches a finite value in the
limit $T\rightarrow0$, in agreement with the Curie-Weiss law for a
free magnetic moment of $S=1/2$. With increasing temperature,
excitations corresponding to higher spins in the trimers are
stimulated and $\chi T$ increases and saturates in high
temperatures. For finite $J_{F}$, $\chi T$ diverges in the limit
$T\rightarrow0$, because of  the formation of ferromagnetic long-range
order in the ground state. Generally, $\chi T$ decreases
monotonically with temperature for ferromagnets, whereas it increases
monotonically with temperature for antiferromagnets. The behavior
of $\chi T$ for this ferrimagnet reflects the interplay of these
two opposite trends. Gapless ferromagnetic excitations reduce the 
magnetization at low $T$, while thermally activated antiferromagnetic 
excitations give rise to the increase of $\chi T$ at high $T$. 
The presence of a minimum in $\chi T$ for all nonzero $J_{F}$ is a
typical feature of one-dimensional (1D) ferrimagnets. Similar behavior was observed
in the alternating $\left(S,s\right)=\left(1,1/2\right)$ spin
chain.\cite{yamamoto98} This minimum moves slightly toward higher
temperature with increasing $\mid J_{F}\mid/J_{AF}$, resulting
from the enhancement of ferromagnetic correlations.

\begin{figure}
\includegraphics[width=7.5cm]{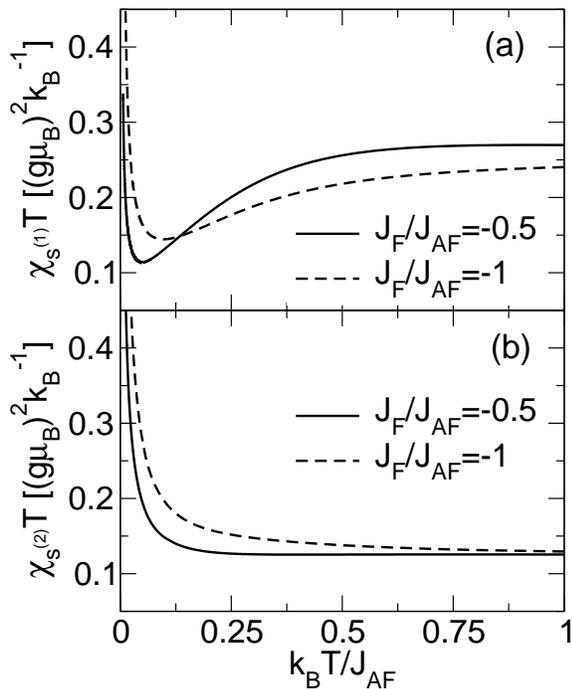}
\caption{\label{stagger}The zero-field staggered susceptibilities
multiplied by temperature. $\chi_{s}^{(1)}$ and $\chi_{s}^{(2)}$
are defined by Eqs. (\ref{eq:staga}) and (\ref{eq:stagb}),
respectively.}
\end{figure}

Figure \ref{stagger} shows the temperature dependence of the staggered
susceptibility. To calculate the staggered susceptibility, we add
a staggered magnetic field $h_{s}$ instead of a uniform field to
the Hamiltonian. Two kinds of staggered fields are considered. One
is a staggered field applied to the whole lattice. The corresponding
Zeeman interaction, staggered magnetization, and susceptibility are
defined by
\begin{eqnarray}
H_s^{(1)} & = & -g\mu_{B}h_{s}\sum_{i=1}^{2L} \left(S_{2i-1}^{z}
-S_{2i}^{z}\right), \\
M_{s}^{(1)} & =&\frac{g\mu_{B}}{4L}\sum_{i=1}^{2L} \langle
S_{2i-1}^{z}-S_{2i}^{z}\rangle,\label{eq:maga}  \\
\chi_{s}^{(1)} & =&\frac{ \partial M_{s}^{(1)}}{\partial
h_{s}}.\label{eq:staga}
\end{eqnarray}
The staggered susceptibility $\chi_{s}^{(1)}$ as defined
corresponds to the spin structure factor at $q=\pi$. Another kind
of staggered field we consider is that applied only to the first and
third spins in each unit cell. The corresponding Zeeman interaction,
staggered magnetization, and susceptibility are defined by
\begin{eqnarray}
H_s^{(2)} & = & -\sum_{i=1}^{L}g\mu_{B}h_{s}
\left(S_{4i-3}^{z}-S_{4i-1}^{z}\right), \\
M_{s}^{(2)} & = & \frac{g\mu_{B}}{4L}\sum_{i=1}^{L}
\langle S_{4i-3}^{z}-S_{4i-1}^{z}\rangle,\label{eq:magb}\\
\chi_{s}^{(2)} & = & \frac{\partial M_{s}^{(2)}}{\partial
h_{s}}.\label{eq:stagb}
\end{eqnarray}
$\chi_{s}^{(2)}$ corresponds to a sum of the spin structure factor
at $q=\pm\pi/2$. As revealed by Fig. \ref{stagger}, both
staggered susceptibilities, $M_s^{(1)}$ and $M_s^{(2)}$, diverge
at $T=0$ in the zero-field limit.

The divergence of the uniform susceptibility $\chi$ as well as the
two staggered susceptibilities $\chi_{s}^{(1)}$ and
$\chi_{s}^{(2)}$ is an indication of the formation of
ferromagnetic as well as antiferromagnetic long-range orders at
zero temperature. This is consistent with a rigorous theorem about
the ferromagnetic and antiferromagnetic orders in the ground state
of Type-II ferrimagnets.\cite{Tian}

\begin{figure}
\includegraphics[width=7.5cm]{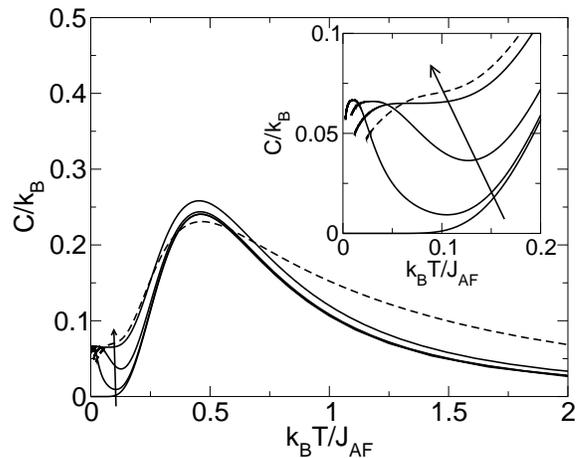}
\caption{\label{ratio}TMRG results of the zero-field specific
heat at $\mid J_{F}\mid/J_{AF}=0,0.05,0.15,0.5,2.0$ in ascending
order along the direction of arrows. The case $\mid
J_{F}\mid/J_{AF}=2.0$ is plotted as a dashed line for visual effect.
The inset shows the low-temperature part of the specific heat.}
\vspace{0.5cm}
\end{figure}

\begin{figure}
\includegraphics[width=7.5cm]{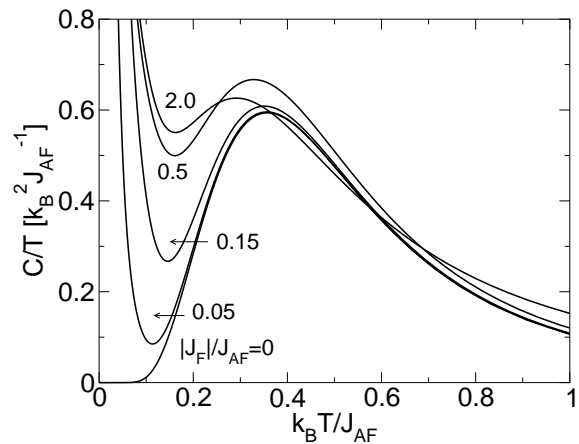}
\caption{\label{ratio3overT}The zero-field specific heat
coefficient $C/T$ versus $T$ for $\mid
J_{F}\mid/J_{AF}=0,\;0.05,\;0.15,\;0.5,\;2.0$.}
\end{figure}

Figure \ref{ratio} shows the temperature dependence of the
zero-field specific heat at different $J_{F}/J_{AF}$. When
$J_{F}=0$, the specific heat drops quickly in low temperatures
because of the finite-energy gap between the ground state and
excitation states. At finite $J_{F}$, a small peak develops in low
temperatures and $C$ shows a small peak-valley structure. The low-temperature
peak of $C$ results from the gapless ferromagnetic
excitations, as illustrated by Fig. \ref{ratio3overT}. In low
temperatures, the temperature dependence of $C/T$ measures the
energy dependence of the density of states of low-lying
ferromagnetic excitations. For finite $J_{F}$, since the density
of states of ferromagnetic excitations diverges at zero energy,
$C/T$ is expected to diverge at $T=0\mbox{ K}$. For $\mid
J_{F}\mid/J_{AF}\gtrsim0.5$, as shown by Fig. \ref{ratio}, the
small peak-valley in the low-temperature specific heat changes
gradually into a hump structure. A more conspicuous peak around
$k_{B}T\sim0.5J_{AF}$ exists in all the cases studied. This
broadened peak, often called the Schottky-like peak, is associated
with gapped antiferromagnetic excitations, as predicted by various
methods\cite{Pati,brehmer,yamamoto98,wu} for the Lieb-Mattis-type
ferrimagnets.

The double-peak structure of the specific heat for ferrimagnets
was also observed by Nakanishi and Yamamoto\cite{Nakanishi} using
modified spin-wave theory. They found that this double-peak
structure of the zero-field specific heat is an intrinsic feature
with topological origin, since the dual features of ferromagnetism
and antiferromagnetism in ferrimagnets can potentially induce a
low-temperature peak as well as an intermediate-temperature peak.
They calculated the low-lying spectra of the tetrameric chain.
Three bands of ferromagnetic excitations with one gapless and two
gapped, and one band of gapped antiferromagnetic excitation were
found. From the above qualitative analysis on the F-F-AF-AF model,
we believe that in the region $\mid J_{F}\mid\ll J_{AF}$, the
low-energy physics of the tetrameric chain is governed by the
ferromagnetic coupling. In this region, the peak should move to
higher temperature with increasing $\mid J_{F}\mid$. Our numerical
results support the above arguments. The positions and the heights
of the low-temperature peaks in the calculations are consistent
with the results for the $S=1/2$ Heisenberg model with the same
ferromagnetic interaction $J_{F}$. With increasing $\mid
J_{F}\mid$, we find that the peak moves toward higher temperature
and eventually mixes with other excitations, leading to a hump
structure. However, the position and the height of the
Schottky-like peak are not sensitive to the ratio $\mid
J_{F}\mid/J_{AF}$. This is different to the modified-spin-wave
theory.\cite{Nakanishi} It should be emphasized that the intrinsic
double-peak structure of the specific heat is not a distinctive
character of bond-alternating-type ferrimagnets. As pointed out by
Nakanishi and Yamamoto, \cite{Nakanishi} this structure appears
when the antiferromagnetic gap is larger than the ferromagnetic
bandwidth. For the $\left(S,s\right)$ alternating-spin-type
ferrimagnets, this condition is satisfied for $S\gg s$.

\begin{figure}
\includegraphics[width=7.5cm]{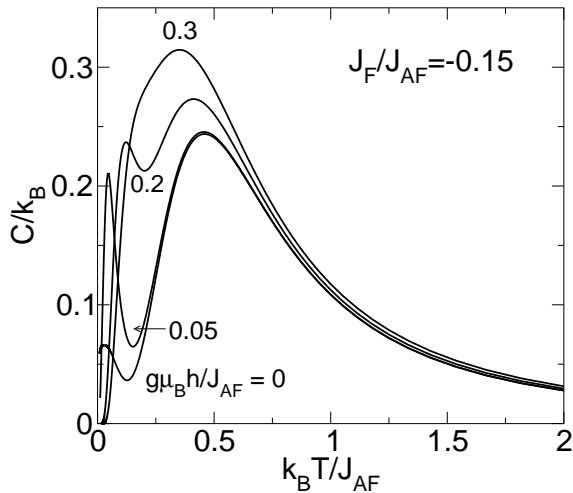}
\caption{\label{0.15}Temperature dependence of the specific
heat for $J_{F}/J_{AF}=-0.15$ at different external magnetic
fields.}
\end{figure}

At finite magnetic field, due to the field-induced splitting of
ferromagnetic excitations, a double-peak structure of the specific
heat can also appear in some ferrimagnets, such as for the case.
$(S,s)=(1,1/2)$ \cite{Maisinger} However, these two cases should
be clearly distinguished. Figure \ref{0.15} shows the temperature
dependence of the specific heat for $J_{F}/J_{AF}=-0.15$ at
different external fields. When a weak field is applied, $C$ shows
a sharp low-temperature peak. With increasing field, the
double-peak structure is smeared out and merges together, due to
the Zeeman splitting.\cite{Maisinger} The Zeeman splitting of the
eigenvalues leads to a finite energy gap in the ferromagnetic
excitations proportional to $h$, but reduces the energy gap in the
antiferromagnetic excitations.

Recently, Stre$\check{\textrm{c}}$ka \emph{et al.} \cite{Strecka1,Strecka2}
investigated thermodynamic properties of a spin-$\frac{1}{2}$ Ising-Heisenberg
chain with F-F-AF-AF bond-alternating interaction using a mapping-transformation
technique. They considered the Ising-type ferromagnetic interaction. 
Nevertheless, their results agree qualitatively with
ours.

\section{Comparison to Experiments\label{sec:Comparison-With-Experiments}}

In this part we calculate the susceptibility and specific heat of
the tetrameric model in the parameter regime relevant to CCPA and
compare the numerical results to the experimental data published
in Refs. \onlinecite{Hagiwara1} and \onlinecite{Hagiwara2}.

\subsection{Magnetization and susceptibility}

\begin{figure}
\includegraphics[width=7.5cm]{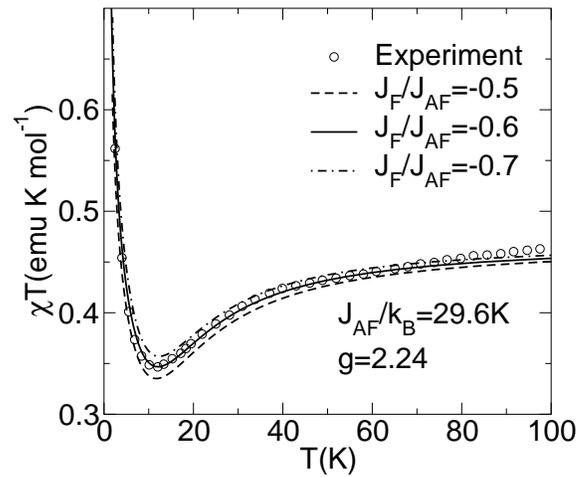}
\caption{\label{sus}Comparison between our numerical results for
the magnetic susceptibility multiplied by temperature and the
experimental data published in Ref. \onlinecite{Hagiwara1}.}
\end{figure}

\begin{figure}
\includegraphics[width=7.5cm]{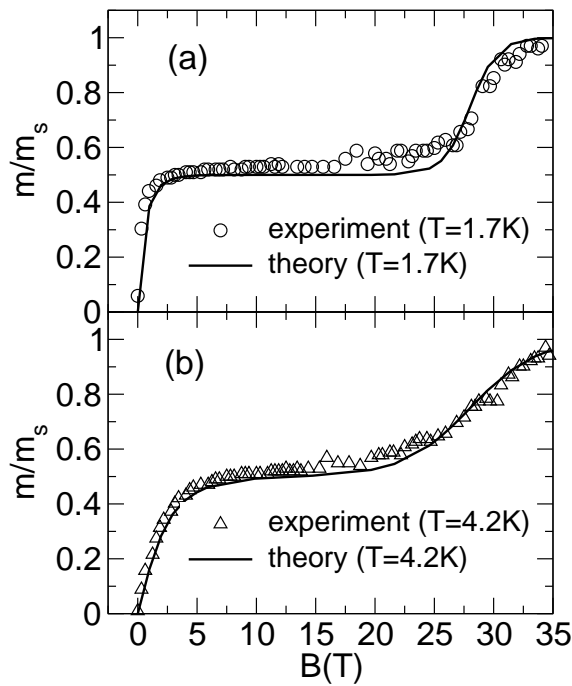}
\caption{\label{mag}Comparison of the measured experimental data
of the field dependence of low-temperature longitudinal
magnetization\cite{Hagiwara2} to the TMRG results. The
parameters used are $J_{AF}/k_{B}=29.6\mbox{ K}$, $J_{F}/J_{AF}=-0.6$,
$g=2.24$. $m_{s}$ is the saturation magnetization.}
\end{figure}

Figure \ref{sus} compares the TMRG results of $\chi(T)T$ for
different $J_{F}/J_{AF}$ to the experimental data for CCPA. The
parameters used are the same as in Ref. \onlinecite{Hagiwara1}. A
good agreement between experiments and calculations is found when
$J_{F}/J_{AF}\approx -0.6$, same as in Ref.
\onlinecite{Hagiwara1}. However, above $70\mbox{ K}$, the numerical curves
are slightly below the experimental ones.

Figure \ref{mag} compares the TMRG results of the field dependence
of the longitudinal magnetization to the experimental data.
Within experimental errors, the TMRG results agree with the
experiments in the low-field regime. In the high-field regime, the
numerical curve deviates slightly from the experimental data. It
is unknown whether this deviation is because of some unknown effects
or measurement errors. At $T=1.7\mbox{ K}$, there is a broad magnetization
plateau. This plateau is a typical feature of a quantum ferrimagnet,
due to macroscopic magnetization of the ground state. At
relatively high temperature $T=4.2\mbox{ K}$, the plateau region shows a
weak field dependence resulting from thermal fluctuations.

At high fields, the magnetization increases sharply and becomes
saturated for $T=1.7\mbox{ K}$. At $T=4.2\mbox{ K}$, the magnetization curve is
smoothed by thermal fluctuations. In the work of Stre$\check{\textrm{c}}$ka
{\it et al.},\cite{Strecka1,Strecka2} quantum fluctuations are restricted to the
antiferromagnetic trimers since the ferromagnetic coupling is
Ising-like. A stepwise increase of magnetization with $B$ at very
low temperature is expected because of the short correlation
length inherent in this quantum ferrimagnet. Nakanishi and
Yamamoto calculated the spectrum of the F-F-AF-AF Heisenberg model
using the modified-spin wave.\cite{Nakanishi} They found that in
the region $\mid J_{F}\mid/J_{AF}<1$, the antiferromagnetic
excitations are almost dispersionless. It is this dispersionless
antiferromagnetic excitation that enhances the magnetization. The
above comparison suggests that the F-F-AF-AF Heisenberg model can
describe quantitatively the temperature dependence of the
susceptibility and field dependence of the magnetization of
CCPA with $J_{F}/J_{AF}\sim-0.6$.

\subsection{Specific heat}

\begin{figure}
\includegraphics[width=7.5cm]{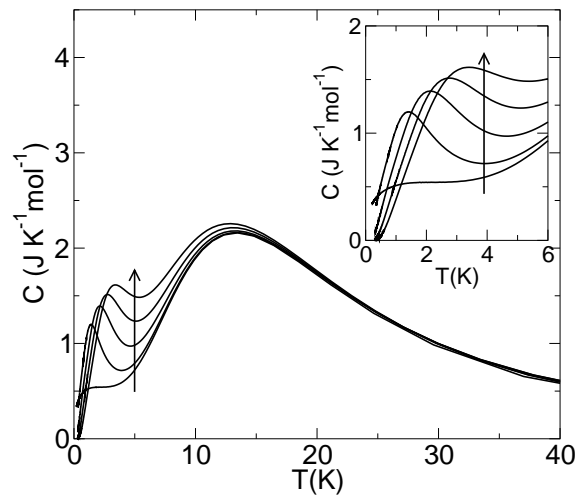}
\caption{\label{heat}TMRG results for the temperature
dependence of the specific heat at $H=0,\;0.5,\;1.0,\;1.5,\;2.0\mbox{ T}$
in ascending order along the direction of arrows. The parameters
used are the same as for Fig. \ref{mag}. The inset shows the low-temperature
part of $C$.}
\vspace{0.5cm}
\end{figure}

Figure \ref{heat} shows the numerical results on the temperature
dependence of the specific heat $C(T)$ at different fields for
$J_{F}/J_{AF}=-0.6$. At zero field, $C$ increases in a certain
power law of $T$ in low temperatures. When an external field is
applied, a double-peak structure appears. With increasing $H$, the
low-temperature peak moves toward higher temperature, but the
high-temperature peak moves in the opposite direction.

\begin{figure}
\includegraphics[width=7.5cm]{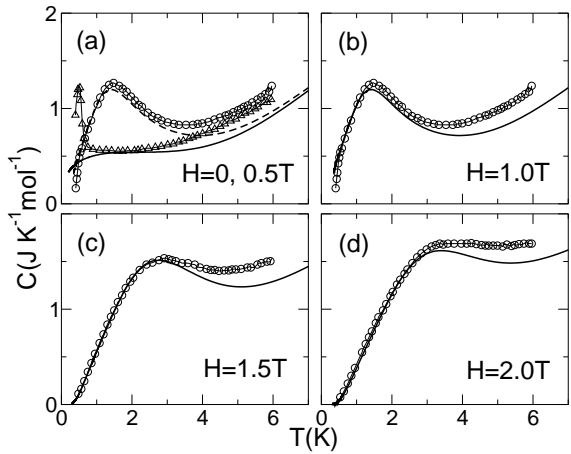}
\caption{\label{heat_comp0}Comparison of the specific heat of CCPA
to the TMRG results. In (a), the triangle (solid), circle
(dashed) lines represent the experimental (numerical) data for
$H=0,\;0.5\mbox{ T}$, respectively; In (b), (c), (d), the circle and
solid lines represent experimental and numerical results,
respectively. The parameters used are the same as for Fig.
\ref{mag}.}
\vspace{0.5cm}
\end{figure}

\begin{figure}
\includegraphics[width=7.5cm]{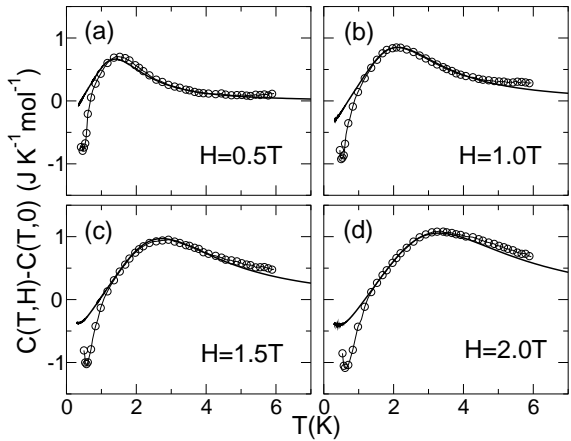}
\caption{\label{heat_comp}Comparison of the specific heat of CCPA
to the TMRG results after subtracting the corresponding
zero-field data. Circles and solid lines represent experimental
and numerical results, respectively. The parameters are the same
as for Fig. \ref{mag}.}
\vspace{0.5cm}
\end{figure}

\begin{figure}
\includegraphics[width=7.5cm]{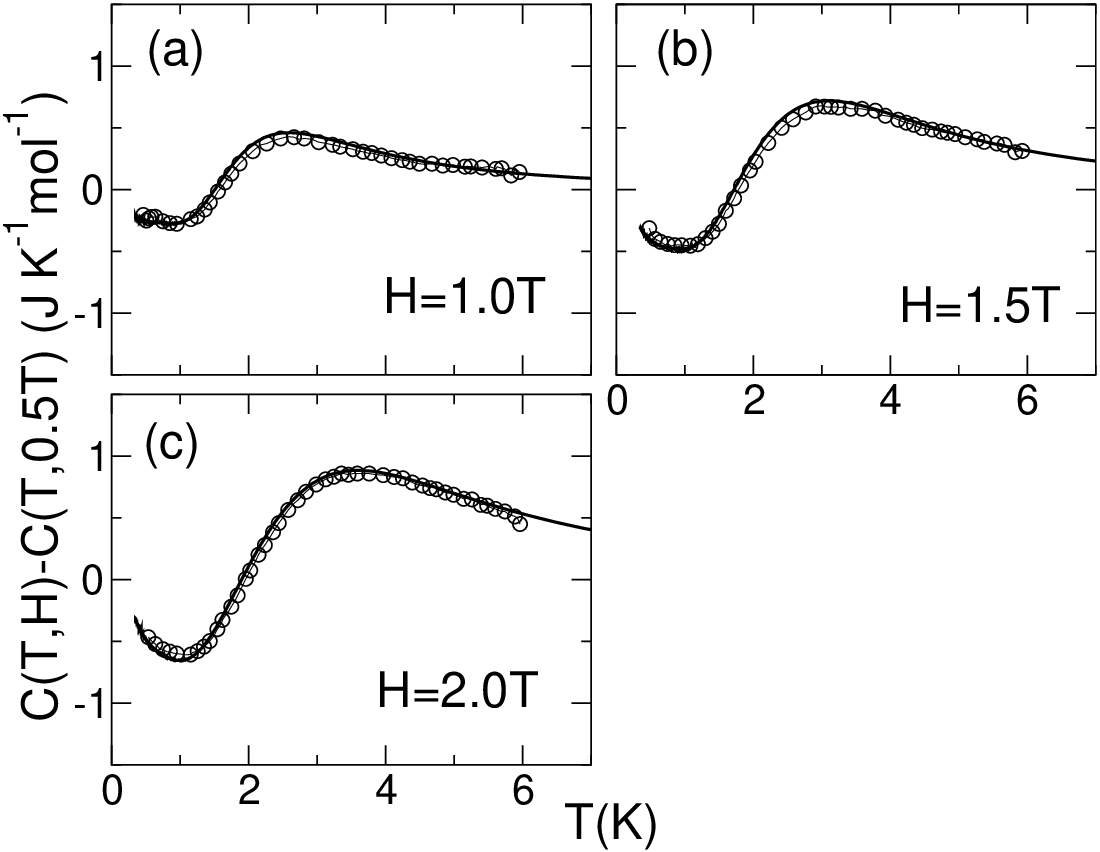}
\caption{\label{heat_comp2}Comparison of the specific heat of CCPA
to TMRG results after subtracting the corresponding data at
$H=0.5\mbox{ T}$. Circles and solid lines represent experimental and
numerical results, respectively. The parameters are the same as
for Fig. \ref{mag}.}
\end{figure}

Figure \ref{heat_comp0} compares the experimental data of $C$ to
our numerical calculations. At finite fields, the numerical
results agree well with the experimental data below $3\mbox{ K}$.
However, in relatively higher temperatures, the experimental
curves are clearly above the numerical ones.

The experimental data shown in Fig. \ref{heat_comp0} were obtained
from the measurement raw data by subtracting the $T^{3}$ phonon
contribution to the specific heat. This subtraction may not always be
accurate. Thus we have reason to suspect that the deviation
at high temperatures may come from the error in the subtraction of
the phonon contribution.

In order to reduce the error in the data subtraction, we compare
our numerical results of the difference of the specific heat at a
finite field and that at zero field, $\Delta C_{1}=C(T,H)-C(T,0)$,
to the corresponding experimental data. As shown in Fig.
\ref{heat_comp}, the agreement between the experimental and
numerical results is improved, especially in high $T$, from $1.3\mbox{ K}$
to $6\mbox{ K}$. However, in low temperatures, the agreement became
worse. This is clearly due to the presence of the experimental sharp peak
around $0.5\mbox{ K}$ at zero magnetic field.

Hagiwara \emph{et al.}\cite{Hagiwara1} suggested that when a
Heisenberg spin chain has more than two kinds of exchange
interactions, the specific heat is expected to show more than one
peak. Manaka \emph{et al.}\cite{Manaka} also found the double-peak
structure in the temperature dependence of the specific heat by
the birefringence measurements in ferromagnetic-dominant F-AF
alternating Heisenberg chains
(CH$_{3}$)$_{2}$CHNH$_{3}$CuCl$_{3}$. From Sec.
\ref{sec:Main-Numerical-Results}, we know that the intrinsic
double-peak structure of the specific heat for the Heisenberg
F-F-AF-AF model could be observed when $\mid J_{F}\mid/J_{AF}$ is
small enough. But for the given parameter $J_{F}/J_{AF}=-0.6$, the
sharp peak at $0.5\mbox{ K}$ is not expected. Stre$\check{\textrm{c}}$ka
\emph{et al.} \cite{Strecka2} also made a comparison of theoretical
results on the F-F-AF-AF Ising-Heisenberg model with experiments.
In order to fit the striking low-temperature peak around $0.5\mbox{ K}$,
$\mid J_{F}\mid/J_{AF}$ had to be drastically reduced, and the
results merely qualitatively agreed with experiments. On the other
hand, from the experimental data, the experimental sharp peak around $0.5\mbox{ K}$
seems to be suppressed by magnetic field [Fig. 4 in Ref.
\onlinecite{Hagiwara1}]. The behavior is different from the usual
situation, as shown in Fig. \ref{0.15}, where the low-temperature
peak of the double-peak structure is strengthened by a weak
external field.

We believe that the experimental sharp peak of the specific heat around $0.5\mbox{ K}$
at zero magnetic field is not an intrinsic property of the model.
It may result from some extrinsic properties of the material, such
as defects or boundary states. Further measurements and more
extensive theoretical and numerical analysis are needed for fully
understanding this behavior.

To avoid the contribution from the experimental sharp low-temperature peak of
the specific heat at zero field, we show in Fig. \ref{heat_comp2}
the results for $\Delta C_{2}=C(T,H)-C(T,0.5\mbox{ T})$, which
subtracts the data in the field $H=0.5\mbox{ T}$, instead of at zero
field. We find that our numerical results agree excellently with
the experiments. This indicates that the experimental sharp low-temperature
peak of the zero-field specific heat is indeed extrinsic.

\section{Conclusion}

We have investigated numerically thermodynamic properties of the
bond-alternating F-F-AF-AF tetramer Heisenberg spin chain using
the TMRG. The temperature dependence of the spin susceptibility
and the specific heat is found to be determined by the competition
of gapless ferromagnetic excitations and gapped antiferromagnetic
excitations. This leads to a crossover in the behavior of $C$ and
$\chi$. With increasing temperature, $\chi T$ drops sharply in low
temperatures, but increases and becomes saturated in high
temperatures. The minimum of $\chi T$ increases gradually with
increasing $|J_F/J_{AF}|$. Both the uniform and staggered
susceptibilities diverge in the zero temperature limit. This
suggests that both ferromagnetic and antiferromagnetic spin
correlations are long-range ordered in the ground state. $C$ shows
a small peak in low temperatures and a stronger and broadened
Schottky-like peak in an intermediate temperature regime. The
nonmonotonic $T$ dependence of $\chi T$ and the double-peak
structure of $C$ are also the characteristic features of
alternating spin ferrimagnets.

Our numerical results for the temperature dependence of $\chi$ and
the difference of $C$ between two different fields $C(H) -
C(0.5\mbox{ T})$ with $J_F/J_{AF} = -0.6$ agree excellently with the
experimental data for CCPA. The field dependence of the
magnetization agrees also with the experiment. We argue that the
sharp peak observed at $\sim 0.5\mbox{ K}$ in the zero-field specific heat
is an extrinsic feature of CCPA. It is likely to be the
contribution of defects or magnetic impurities. Further
experimental measurements with high-quality samples are desired to
clarify this issue.
\\

\begin{acknowledgments}
This work was supported by the National Natural Science Foundation of China.
Part of the numerical work of this project was performed on the HP-SC45 Sigma-X
parallel computer of ITP and ICTS, CAS.
\end{acknowledgments}


\begin{thebibliography}{10}

\bibitem{Lieb}E. Lieb and D. Mattis, J. Math. Phys. \textbf{3}, 749 (1962).

\bibitem{Pati}S. K. Pati, S. Ramasesha, and D. Sen, Phys. Rev. B \textbf{55}, 8894
(1997).

\bibitem{Yamamoto98prb}S. Yamamoto, S. Brehmer and H.-J. Mikeska, Phys. Rev. B \textbf{57},
13610 (1998).

\bibitem{yamamoto98}S. Yamamoto, T. Fukui, K. Maisinger, and U. Shollw$\ddot{\mathrm{o}}$ck,
J. of Phys. : Cond. Mat. \textbf{10}, 11033 (1998).

\bibitem{wu}C. Wu, B. Chen, X. Dai, Y. Yu, and Z.-B. Su, Phys. Rev. B \textbf{60},
1057 (1999).

\bibitem{Manaka}H. Manaka, I. Yamada, T. Kikuchi, K. Morishita, and K. Iio, J. Phys.
Soc. Jpn. \textbf{70}, 2509 (2001).

\bibitem{Hagiwara1}M. Hagiwara, K. Minami, and H. A. Katori, Prog. Theor. Phys. Suppl.
\textbf{145}, 150 (2002).

\bibitem{Hagiwara2}M. Hagiwara, Y. Narumi, K. Minami, K. Kindo, H. Kitazawa, H. Suzuki,
N. Tsujii, and H. Abe, J. Phys. Soc. Jpn. \textbf{72}, 943 (2003).

\bibitem{Inorg}A. Escuer, R. Vicente, M. S. El Fallah, M. A. S. Goher, and F. A.
Mautner, Inorg. Chem. \textbf{37}, 4466 (1998).

\bibitem{yamamoto04}S. Yamamoto, Phys. Rev. B \textbf{69}, 064426 (2004).

\bibitem{Nakanishi}T. Nakanishi and S. Yamamoto, Phys. Rev. B \textbf{65}, 214418 (2002).

\bibitem{kahn}O. Kahn, Struct. Bonding (Berlin) \textbf{68}, 89 (1987); O. Kahn,
Y. Pei, and Y. Journaux, in \emph{Inorganic Materials}, edited by
D. W. Bruce and D. O' Hare (Wiley, New York, 1995), p.25.

\bibitem{White}S. R. White, Phys. Rev. Lett. \textbf{69}, 2863 (1992).

\bibitem{brehmer}S. Brehmer, H.-J. Mikeska, and S. Yamamoto, J. Phys.: Condens. Mat.
\textbf{9}, 3921 (1997).

\bibitem{Bursill} R. J. Bursill, T. Xiang, and G. A. Gehring, J. Phys. Condens. Matter
\textbf{8}, L583 (1996).

\bibitem{Wang} X. Wang and T. Xiang, Phys. Rev. B \textbf{56}, 5061 (1997).

\bibitem{Xiang} T. Xiang and X. Wang, in \emph{Density-Matrix Renormalization:
A New Numerical Method in Physics}, edited by I.
Peschel, X. Wang, M. Kaulke and K. Hallberg (Springer, New York, 1999), pp. 149-172.

\bibitem{Tian}Guangshan Tian, Phys. Rev. B \textbf{56}, 5355 (1997).

\bibitem{Maisinger}K. Maisinger, U. Schollw$\ddot{\mathrm{\textrm{o}}}$ck, S. Brehmer,
H. J. Mikeska, and S. Yamamoto, Phys. Rev. B \textbf{58}, R5908 (1998).

\bibitem{Strecka1}J. Stre$\check{\textrm{c}}$ka, M. Ja$\check{\textrm{s}}$$\check{\textrm{c}}$ur,
M. Hagiwara, and K. Minami, Czech. J. Phys. \textbf{54}, D583 (2004).

\bibitem{Strecka2}J. Stre$\check{\textrm{c}}$ka, M. Ja$\check{\textrm{s}}$$\check{\textrm{c}}$ur,
M. Hagiwara, Y. Narumi, K. Kindo, and K. Minami, cond-mat/0406680 (unpublished).

\end{thebibliography}
\end{document}